\documentclass[pdfusetitle,aps,reprint,prd,twocolumn,superscriptaddress,nofootinbib]{revtex4-1}

\usepackage{amssymb}
\usepackage{amsmath}
\usepackage{physics}
\usepackage{bbm}
\usepackage{epsfig}
\usepackage{breakurl}
\usepackage{float}
\usepackage{multirow}
\usepackage{array}
\usepackage{bm}
\usepackage{color}
\usepackage{tikz}
\usepackage[utf8]{inputenc}
\usepackage{braket}
\usepackage{graphicx}
\usepackage{footmisc}
\usepackage{tensor}
\usepackage[normalem]{ulem}
\usepackage{dblfloatfix}
\usepackage[T1]{fontenc}
\usepackage{array}
\usepackage{booktabs}
\usepackage{multirow}
\usepackage{amstext}
\usepackage{multirow}

\usepackage[colorlinks=true,citecolor=blue,linkcolor=blue,urlcolor=blue]{hyperref}
\usepackage{tikz}
\usetikzlibrary{arrows.meta}

\newcommand{\A}{\mathcal A}
\newcommand{\HH}{\mathcal H}
\newcommand{\PP}{\mathbb P}

\newcommand{\tdelay}{\Delta T}
\newcommand{\diff}{\mathrm{diff}}

\newcommand{\bfone}{\mathbf 1}
\begin{document}

\title{Causality and the Equivalence Principle for Higher Energy Scattering}

\author{Yu-tin Huang}
\affiliation{Department of Physics and Center for Theoretical Physics, National
Taiwan University, Taipei 10617, Taiwan}
\affiliation{Physics Division, National Center for Theoretical Sciences, Taipei
10617, Taiwan}
\affiliation{Max Planck{-}IAS{-}NTU Center for Particle Physics, Cosmology and
Geometry, Taipei 10617, Taiwan}

\author{Lukas W. Lindwasser}
\affiliation{Department of Physics and Center for Theoretical Physics, National
Taiwan University, Taipei 10617, Taiwan}

\date{\today}

\begin{abstract}
Recently, it was proposed that the leading high-energy behavior of scattering amplitudes is universal, independent of charge, thereby extending the equivalence principle beyond the graviton pole. In this Letter, we derive a sharper causality constraint on such behavior by studying the Regge limit of colored scattering. Parameterizing a trajectory by $s^{\alpha(t)}$ with $\alpha(0)=2-\delta$, we analyze the Shapiro/Wigner--Smith time-delays in the irreducible scattering channels. We show that any non-singlet trajectory with $\delta< 1/2$ produces a growing sign-indefinite time-delay (with $\delta=1/2$ a marginal, dimension-dependent case), which becomes dominant in the Regge diffusion region in the weak-gravity regime. The essential point is that, while the eikonal phase is naturally organized in $t$-channel irreducible representations, the physical time-delays are its eigenvalues in the $s$-channel. A non-singlet exchange therefore recouples into the physical channels with both signs, inevitably producing a negative time-delay in at least one channel.

\end{abstract}

\maketitle

\section{Introduction}
The equivalence principle admits a sharp formulation in the language of scattering amplitudes. At four points, elastic scattering of two species $a,b$ contains a universal graviton pole whose residue factorizes into two matter-matter-graviton three-point amplitudes whose gravitational coupling $\kappa$ is universal due to Weinberg’s soft graviton theorem \cite{Weinberg:1965nx}. Consequently, the forward amplitude contains a universal contribution $\mathcal A_4^{ab}\sim \kappa^2 s^2/(-t)$, independent of flavor, charge, or other internal labels. In this sense, the equivalence principle appears at four points as the universal residue of the $t$-channel graviton pole.

Recently, Calisto, Cheung, Remmen, Sciotti, and Tarquini proposed a high-energy extension of this principle \cite{Calisto:2026ep}. In particular, the universality of the massless spin-2 pole is extended to the leading high energy behavior, large $s$, at fixed $t$. Charge dependent contributions may be present, but must remain subleading relative to the universal leading behavior. Parameterizing the $s\rightarrow \infty$ and fixed $t<0$ limit of the amplitude as $s^{\alpha(t)}$, with the intercept $\alpha(0)=2-\delta$, the statement is then $\delta$ is a finite constant for any non-singlet channels.  This raises the natural question of whether such a strengthened equivalence principle follows from general consistency conditions. 

The key diagnostic is the Shapiro/Wigner--Smith time-delay \cite{Shapiro:1964uw}. In high-energy scattering, the Regge amplitude determines an eikonal phase in impact-parameter space, and physical time-delays are obtained from its eigenvalues. A resolvable time-advance signals an obstruction to causal UV completion. This logic underlies the analysis by Camanho, Edelstein, Maldacena, and Zhiboedov \cite{Camanho:2014apa}, who showed that higher-derivative corrections to graviton interactions can produce negative time-delays (for discussion on the resolvability of such causality violation within gravitational EFT framework, see~\cite{deRham:2020zyh, Chen:2021bvg}). 

In this Letter, we apply the same causality criterion to the internal charge structure of the Regge amplitude. We consider elastic scattering in the Regge regime and decompose the amplitude into irreducible exchange channels. The eikonal phase is then an operator acting on the incoming charge space, and the physical time-delays are its eigenvalues. While the graviton singlet exchange gives a universal positive contribution, it can be consistently suppressed in the weak-gravity regime, defined through the dimensionless parameter $\lambda_{\rm grav}\equiv G_N M_\ast^{D-2}\ll1$, where $M_\ast$ is the Regge scale. In such a regime, we demonstrate that any non-singlet trajectory with $\delta<\frac{1}{2}$ will invariably lead to a negative time-delay eigenvalue in some incoming
channel.

\section{The Eikonal phase and time-delay}

Let $t=-q^2$ and $d_\perp=D-2$.  The leading contribution to the eikonal phase is the transverse Fourier transform of the four-point amplitude,
\begin{equation}
  \chi(s,b)= \frac{1}{2s}
  \int \frac{\mathrm{d}^{d_\perp}q}{ (2\pi)^{d_\perp}}
  e^{\mathrm{i}q\cdot b}\,\A(s,-q^2) .
  \label{eq:eikonal}
\end{equation}
In the elastic eikonal regime, $S_{\text{eik}}(s,b)=\exp[\mathrm{i}\chi(s,b)]$.  The corresponding time-delay is
\begin{equation}
  \tdelay(s,b)=\frac{\partial}{\partial E}\,\mathrm{Re}\,\chi(s,b),
  \qquad s\simeq 4E^2 .
  \label{eq:delaydef}
\end{equation}
A negative eigenvalue larger than the wavepacket resolution is a resolvable time-advance and is incompatible with asymptotic causality~\cite{Wigner,Smith, Martin:1976timeDelay, Camanho:2014apa}.

We now consider the standard Regge parametrization of the high energy fixed $t$ scattering,
\begin{equation}
  \A_\rho(s,t)\simeq \eta_\rho(t)\,\beta_\rho(t)
  \left(\frac{s}{ M_*^2}\right)^{\alpha_\rho(t)} .
  \label{eq:reggeform}
\end{equation}
Here the label $\rho$ denotes the
$t$-channel quantum numbers; for a charge-neutral amplitude it may be omitted.
This form is the usual Regge-pole description of high-energy scattering at
small momentum transfer~\cite{Collins}, and is also the form underlying the
impact-parameter analysis of trans-Planckian and gravitational Regge
scattering~\cite{ACV,Haring:2022sdp}. The trajectory $\alpha_\rho(t)$ fixes the
energy growth.  We will write
\begin{equation}
  \alpha_\rho(-q^2)=\alpha_\rho(0)-\alpha'_\rho q^2+\cdots,
  \label{eq:trajectoryexp}
\end{equation}
so that the intercept $\alpha_\rho(0)$ controls the overall power of $s$, while
$\alpha'_\rho$ controls transverse Regge diffusion in impact-parameter space. A pole in $\beta_\rho(t)$ will require the intercept to be integer, reflecting the exchange of massless spin-$\alpha(0)$ state. Since a massless spin-2 particle must be a singlet, the question at hand is: \textit{if $\beta_\rho(t)$ is regular near $t=0$, can $\alpha_\rho(0)$ asymptote to $2$ as $t\rightarrow 0^-$}?

The factor $\eta_\rho(t)$ is the signature factor, or Regge phase.\footnote{For a pole
of definite signature $\tau=\pm1$, one common convention is
\begin{equation}
  \eta_{\tau}(t)= -\,{1+\tau e^{-\mathrm{i}\pi\alpha_\rho(t)}\over
  \sin\pi\alpha_\rho(t)} ,
  \label{eq:signaturefactor}
\end{equation}
up to convention-dependent phases and normalization. Its role is to implement
the correct crossing properties and to determine the complex phase of the
Regge exchange.} Since the time-delay depends on
$\partial_E\mathrm{Re}\,\chi$, the relevant coefficient below is the real part
of the product $\eta_\rho(t)\beta_\rho(t)$. At large impact parameter, there are two regions of interest.  Define,
\begin{equation}
  \alpha_\rho(0)=2-\delta_\rho,
  \qquad
  L\equiv \log\left( {s\over M_*^2}\right),
  \qquad
  b_{\diff,\rho}^2\equiv \alpha'_\rho L .
  \label{eq:interceptdiff}
\end{equation}
and expand the real part of the signature-residue product near $t=0$ as
\begin{equation}
  {1\over 2M_*^2}\,\mathrm{Re}\!\big[\eta_\rho(-q^2)\beta_\rho(-q^2)\big]
  ={c_{\rho,-1}\over q^2}+c_{\rho,0}+O(q^2).
  \label{eq:cbetaexp}
\end{equation}
The coefficient $c_{\rho,-1}$ is present only when the $t$-channel exchange contains a massless pole.  The coefficient $c_{\rho,0}$ is the leading term when the residue is regular. At large impact parameter the integral is well approximated by
\begin{equation}
\begin{split}
  \mathrm{Re}\,\chi_\rho(s,b)
  \simeq{}& \left({s\over M_*^2}\right)^{1-\delta_\rho}
  \Big[c_{\rho,-1}P_\rho(b,L)
  \\
  &\qquad\qquad +c_{\rho,0}D_\rho(b,L)+\cdots\Big],
\end{split}
  \label{eq:phasekernels}
\end{equation}
where $P_\rho(b,L)$ is the pole contribution 
\begin{equation}
  P_\rho(b,L)=\int_{\alpha'_\rho L}^{\infty}\!\mathrm{d} u\,
  (4\pi u)^{-d_\perp/2}e^{-b^2/4u}.
  \label{eq:Pkernel}
\end{equation}
and $D_\rho(b,L)$ is the Fourier transform of the Gaussian approximation: 
\begin{equation}
  D_\rho(b,L)= (4\pi\alpha'_\rho L)^{-d_\perp/2}
  \exp\!\left[-{b^2\over 4\alpha'_\rho L}\right]
  \label{eq:Dkernel}.
\end{equation}
For $d_\perp>2$ the former interpolates between the transverse propagator
\begin{equation}
 {\rm Tail:} \quad P_\rho(b,L)\simeq G_\perp(b)=\int {\mathrm{d}^{d_\perp}q\over(2\pi)^{d_\perp}}
  {e^{\mathrm{i}q\cdot b}\over q^2}
  \propto {1\over b^{D-4}}
  \label{eq:Gperp}
\end{equation}
in the tail, where $b\gg b_{\diff,\rho}$ and
\begin{equation}
 {\rm Diffusion\, disk:} \quad P_\rho(b,L)\simeq {\alpha'_\rho L\over d_\perp/2-1}\,D_\rho(b,L)
  \label{eq:Pdisk}
\end{equation}
inside the diffusion disk
\begin{equation}
  R_S(\sqrt s)\ll b\lesssim b_{\diff,\rho}(s),
  \qquad
  R_S\sim (G_N\sqrt s)^{\frac{1}{D-3}}\,.
  \label{eq:diskwindow}
\end{equation}
The corresponding time-delay follows by differentiating with respect to $E$, using
$s\simeq4E^2$ 
\begin{equation}
\begin{split}
  \tdelay_\rho(s,b)
  &\simeq {2\over E}\left({s\over M_*^2}\right)^{1-\delta_\rho}
  \Big\{
  c_{\rho,-1}\big[(1-\delta_\rho)P_\rho-\alpha'_\rho D_\rho\big]
  \\
  &\hspace{0.4cm}
  +c_{\rho,0}D_\rho
  \left[(1-\delta_\rho)-{d_\perp\over 2L}
  +{b^2\over 4\alpha'_\rho L^2}\right]
  +\cdots\Big\} .
\end{split}
  \label{eq:generaldelay}
\end{equation}
It is useful to display separately the two impact-parameter regimes.

In the tail, $b\gg b_{\diff,\rho}$, the pole kernel approaches the transverse
propagator while the regular kernel is exponentially suppressed,
\begin{equation}
\begin{split}
  \tdelay^{\rm tail}_\rho(s,b)
  \simeq {2\over E}\left({s\over M_*^2}\right)^{1-\delta_\rho}
  \Bigg\{
  c_{\rho,-1}(1-\delta_\rho)G_\perp(b)
  \\
  +c_{\rho,0}(4\pi\alpha'_\rho L)^{-d_\perp/2}
  e^{-x}
  \left[(1-\delta_\rho)-{d_\perp\over 2L}
  +{x\over L }\right]
  \Bigg\}.
\end{split}
  \label{eq:taildelay}
\end{equation}
where we have defined $x\equiv b^2/(4\alpha'_\rho L)=O(1)$. Thus the time-delay in the asymptotic tail region is controlled by the massless pole, 
\begin{equation}
  |\tdelay^{\rm tail}_{\rho,{\rm pole}}|
  \sim |c_{\rho,-1}|\,
  s^{\frac{1}{2}{-}\delta_\rho}(1-\delta_\rho)\,G_\perp(b)\,,
  \label{eq:tailpolescaling}
\end{equation}
which is only present for gravity. 

Inside the diffusion disk, $R_S\ll b\lesssim b_{\diff,\rho}$, one may
use Eq.~\eqref{eq:Pdisk}.  The delay becomes
\begin{align}
  |\tdelay^{\rm diff}_{\rho,{\rm pole}}|
  &\sim |c_{\rho,-1}|\,
  s^{\frac{1}{2}{-}\delta_\rho}
  (\alpha'_\rho L)^{1{-}\frac{d_\perp}{2}}
  e^{-x}
  \left|{1-\delta_\rho\over \frac{d_\perp}{2}-1}-{1\over L}\right|,
  \nonumber  \\
  |\tdelay^{\rm diff}_{\rho,{\rm reg}}|
  &\sim |c_{\rho,0}|\,
  s^{\frac{1}{2}{-}\delta_\rho}
  (\alpha'_\rho L)^{{-}\frac{d_\perp}{2}}
  e^{-x}
  \left|(1-\delta_\rho)+{x-\frac{d_\perp}{2}\over L}\right|.
  \label{eq:diffregularscaling}
\end{align}
where we have written the two contributions separately.
The power of $s$ is therefore still governed by $s^{\frac{1}{2}-\delta_\rho}$, but the
logarithmic prefactors and the Gaussian profile distinguish the two regions.
For $\delta_\rho<1/2$ the power growth dominates these logarithms throughout
the diffusion disk.  For $\delta_\rho = 1/2$ the delay is  logarithmically suppressed for $d_\perp>2$.

Thus in summary, within the diffusion disk a power-growing time-delay occurs for $\delta_\rho<1/2$, while at $\delta_\rho=1/2$ the delay is marginal: it survives only at $d_\perp=2$ and is logarithmically suppressed for $d_\perp>2$.

\section{Time-delay in charge space}

For colored external particles the amplitude is a matrix in charge space.  The Regge limit is naturally diagonal in $t$-channel projectors,\footnote{If the incoming particles are in different representations, $\PP^{(t)}_{\rho}$ is not a projector in the sense that $\PP^{(t)}_{\rho_1}\cdot\PP^{(t)}_{\rho_2}=\delta_{\rho_1\rho_2}\PP^{(t)}_{\rho_1}$. There is however a privileged set of independent invariant structures $\PP^{(t)}_\rho$, each one associated with an exchanged irreducible representation $\rho$ common to the $t$-channel Hilbert spaces $R_1\otimes \overline{R}_1$ and $R_2\otimes \overline{R}_2$, respectively.}
\begin{equation}
  \A(s,t)=\sum_\rho \A_\rho(s,t)\PP^{(t)}_\rho,
  \qquad
  \chi(s,b)=\sum_\rho \chi_\rho(s,b)\PP^{(t)}_\rho .
  \label{eq:tchannel}
\end{equation}
The physical incoming Hilbert space is $\HH_{\rm in}=R_1\otimes R_2$, whose irreducible components define $s$-channel orthogonal projectors $\PP^{(s)}_\lambda$.  The physical time-delays are then the eigenvalues in this basis. The two appropriately chosen invariant bases are linearly mapped to each other via a real invertible recoupling matrix,
\begin{equation}
  \PP^{(t)}_\rho=\sum_\lambda M^{ts}_{\rho\lambda}\PP^{(s)}_\lambda,
  \label{eq:recouple}
\end{equation}
\begin{figure}
    \centering
    \IfFileExists{Figures/TtoS.tikz}{\begin{tikzpicture}[
  line cap=round,
  line join=round,
  very thick,
  >=Stealth,
  scale=1.0,
  amp/.style={circle, fill=black, inner sep=1.6pt},
  mom/.style={->, shorten >=2pt, shorten <=2pt}
]


\begin{scope}[shift={(0,0)}]
  \coordinate (A) at (0,0.45);
  \coordinate (B) at (0,-0.45);
\draw[mom, -{Stealth[scale=0.7]}] (0.0,-.0) -- (0,-.2);
\node[left] at (-0.65,1.1){$R_1$};
\node[right] at (0.65,1.1){$\overline{R}_1$};
\node[left] at (-0.65,-1.1){$R_2$};
\node[right] at (0.65,-1.1){$\overline{R}_2$};
  \draw (A) -- (B);
  \draw (A) -- (-0.65,1.1);
  \draw (A) -- (0.65,1.1);
  \draw (B) -- (-0.65,-1.1);
  \draw (B) -- (0.65,-1.1);

  \fill (A) circle (1.5pt);
  \fill (B) circle (1.5pt);

  \draw[mom, -{Stealth[scale=0.7]}] (-0.35,.8)--(-0.491421,.94142);
  \draw[mom, -{Stealth[scale=0.7]}] (0.391421,0.84142) -- (0.25,.7);
  \draw[mom, -{Stealth[scale=0.7]}] (-0.391421,-.84142) -- (-0.25,-.7);
  \draw[mom, -{Stealth[scale=0.7]}] (0.38,-.83) -- (0.521421,-.97142);

  \node[right] at (0.12,0) {\small$\rho$};
\end{scope}

\node at (1.,0) {\Large $=$};
\node at (1.8,-.15) {\Large $\displaystyle\sum_{\lambda}$};
\node at (2.7,0) {\large $M^{ts}_{\rho\lambda}$};

\begin{scope}[shift={(4.9,0)}]
  \coordinate (L) at (-.7,0);
  \coordinate (R) at (.7,0);
\node[left] at (-1.3,0.75){$R_1$};
\node[right] at (1.3,0.75){$\overline{R}_1$};
\node[left] at (-1.3,-.75){$R_2$};
\node[right] at (1.3,-.75){$\overline{R}_2$};
  \draw (L) -- (R);
  \draw (L) -- (-1.3,0.75);
  \draw (L) -- (-1.3,-0.75);
  \draw (R) -- (1.3,0.75);
  \draw (R) -- (1.3,-0.75);
\draw[mom, -{Stealth[scale=0.7]}] (-0.0,0) -- (0.2,0);
  \fill (L) circle (1.5pt);
  \fill (R) circle (1.5pt);

  \draw[mom, -{Stealth[scale=0.7]}]  (-.83506,0.156326) -- (-1.15,0.5625);
  \draw[mom, -{Stealth[scale=0.7]}] (-1.,-0.375) -- (-.87506,-0.218826);
  \draw[mom, -{Stealth[scale=0.7]}] (1.,0.375) -- (.87506,0.218826) ;
  \draw[mom, -{Stealth[scale=0.7]}]  (.83506,-0.156326)--(1.15,-0.5625);

  \node[above] at (0,0.) {\small$\lambda$};
\end{scope}

\end{tikzpicture}}{\fbox{\parbox{0.8\linewidth}{\centering\vspace{1.2cm} [birdtrack diagram \texttt{Figures/TtoS.tikz}] \vspace{1.2cm}}}}
    \caption{A diagrammatic representation \cite{Birdtracks} of Eq.~\eqref{eq:recouple}, with the external color representations considered in this Letter, assuming all incoming states.}
    \label{fig:TtoS}
\end{figure}
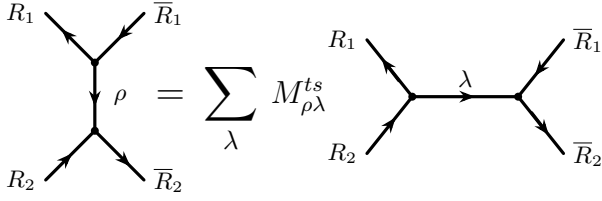
As a consequence, the time-delay for each irreducible representation is a linear combination of derivatives on the eikonal phase
\begin{equation}
  \tdelay_\lambda(s,b)=\sum_\rho M^{ts}_{\rho\lambda}
  {\partial\over\partial E}\,\mathrm{Re}\,\chi_\rho(s,b) .
  \label{eq:delaymatrix}
\end{equation}
Causality requires all of these eigenvalues to be non-negative, up to unresolvable corrections.

As an example, consider the contribution from the graviton pole. In the tail, which is dominated by massless pole, the eikonal phase is simply
\begin{equation}
  \chi_{\rm grav}(s,b)=4\pi G_Ns\,G_\perp(b)\,\PP^{(t)}_{\bfone}
  \label{eq:gravphase_tail}
\end{equation}
where $\PP^{(t)}_{\bfone}$ is the $t$-channel singlet projector. As we will see shortly, in general $M^{ts}_{\bfone\lambda}$ is universal and positive. Hence we find 
\begin{equation}
  \tdelay^{\rm grav}_\lambda>0,
  \qquad \forall \lambda,
  \label{eq:gravpositive}
\end{equation}
with the same positive shift in every incoming representation.

A gauge pole is different.  For a Yang--Mills exchange,
\begin{equation}
  \A_{\rm gauge}(s,-q^2)= {g_{\rm YM}^2s\over q^2}\,\mathsf T_{12} +\cdots ,\; \mathsf T_{12}\equiv \sum_a T_1^a T_2^a 
  \label{eq:gaugeamp}
\end{equation}
where $T_1$ and $T_2$ are adjoint operators acting on the incoming irreducible representations $R_1$ and $R_2$ respectively.
Using the operator identity $T_{\rm tot}^2
  =T_1^2+T_2^2+2\mathsf T_{12}$, we can easily convert the group theory numerator in terms of $s$-channel projectors,
\begin{equation}
  \mathsf T_{12}
  =\sum_{\lambda\subset R_1\otimes R_2}
  {1\over 2}\left[C_2(\lambda)-C_2(R_1)-C_2(R_2)\right]\PP^{(s)}_\lambda \,
  \label{eq:T12ProjectorExpansion}
\end{equation}
where $C_2$ represents the Casimir $T^2$ acting on the appropriate representation. Thus we see that the adjoint exchange is not sign definite in the $s$-channel projector basis. This is nothing but the charge dependent phase that reflects attractive or repulsive dynamics in gauge interactions. There is however no causality violation for the gauge boson exchange, since $\delta_{\rm Adj}=1$ in Eq.~(\ref{eq:diffregularscaling}).

In the following section, we will see such a sign-indefinite property of conversion between $t$- and $s$-channels holds for all irreducible representations, except the singlet.

\section{Non-singlet channels always give negative time-delay}
In the following we will assume that gravity can be taken to be arbitrarily weak, and hence any non-gravitational time-delay can become dominant. 

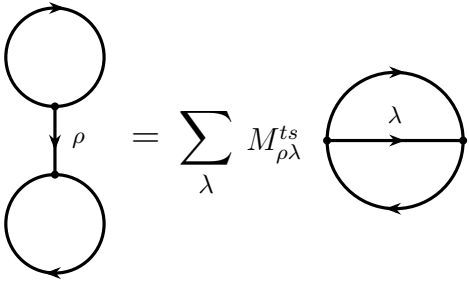
\begin{figure}
    \centering
        \IfFileExists{Figures/ShursLemma.tikz}{\begin{tikzpicture}[
  line cap=round,
  line join=round,
  very thick,
  >=Stealth,
  scale=1.0,
  amp/.style={circle, fill=black, inner sep=1.6pt},
  mom/.style={->, shorten >=2pt, shorten <=2pt}
]

\begin{scope}[shift={(0,-4.2)}]
  \coordinate (T) at (0,0.45);
  \coordinate (B) at (0,-0.45);

  \draw (0,1.1) circle [x radius=0.65, y radius=0.65];
  \draw (0,-1.1) circle [x radius=0.65, y radius=0.65];
  \draw (T) -- (B);
  \fill (T) circle (1.5pt);
  \fill (B) circle (1.5pt);

  \draw[mom, -{Stealth[scale=0.7]}] (-0.0,1.75) -- (0.2,1.75);
  \draw[mom, -{Stealth[scale=0.7]}] (0.0,-1.75) -- (-0.2,-1.75);
  \node[right] at (0.1,00) {\small$\rho$};
  \draw[mom, -{Stealth[scale=0.7]}] (0.0,-.0) -- (0,-.2);
\end{scope}

\node at (1.2,-4.2) {\Large $=$};
\node at (2.,-4.35) {\Large $\displaystyle\sum_{\lambda}$};
\node at (2.9,-4.25) {\large $M^{ts}_{\rho\lambda}$};

\begin{scope}[shift={(4.5,-4.2)}]
  \draw (0,0) circle [radius=.9];
  \coordinate (L) at (-.9,0);
  \coordinate (R) at (.9,0);
 \draw (L) -- (R);
\draw[mom, -{Stealth[scale=0.7]}] (-0.0,0) -- (0.2,0);
  \fill (L) circle (1.5pt);
  \fill (R) circle (1.5pt);
 
  \draw[mom, -{Stealth[scale=0.7]}] (-0.,.9) -- (0.2,0.9);
  \draw[mom, -{Stealth[scale=0.7]}] (0.,-.9) -- (-0.2,-0.9);

  \node[above] at (0,0.1) {\small$\lambda$};
\end{scope}
\end{tikzpicture}}{\fbox{\parbox{0.8\linewidth}{\centering\vspace{1.2cm} [birdtrack diagram \texttt{Figures/ShursLemma.tikz}] \vspace{1.2cm}}}}
    \caption{A diagrammatic representation of the $s$-channel trace of Eq.~\eqref{eq:recouple}. The $s$-channel trace of a non-singlet $t$-channel projector vanishes $\text{tr}_s\PP^{(t)}_\rho=0$ by Schur's lemma, yielding Eq.~\eqref{eq:weightedzerosum}. }
    \label{fig:ShursLemma}
\end{figure}

The $s$-channel projectors are normalized such that its trace adapted to the $\mathcal{H}_{\text{in}}=R_1\otimes R_2$ Hilbert space gives the dimension of the representation
\begin{align}
\text{tr}_s\PP^{(s)}_\lambda =\text{dim}(\lambda)
\end{align}
Taking the $s$-channel trace of Eq.~\eqref{eq:recouple}, the right-hand side becomes a dimension-weighted sum, while the left-hand side, $\text{tr}_s\PP^{(t)}_\rho$, generally vanishes for any non-singlet irreducible representation $\rho$. Consequently the recoupling matrix $M^{ts}_{\rho\lambda}$ has the following property,
\begin{align}
\label{eq:weightedzerosum}
\sum_{\lambda}M^{ts}_{\rho\lambda}\text{dim}(\lambda)\propto\delta_{\rho \bfone}
\end{align}
The details of this statement for a generic compact Lie group is expressed diagrammatically in Fig. \ref{fig:ShursLemma}, and discussed in App. \ref{app:A}.

Furthermore, as explained in App. \ref{app:B} the singlet row $M^{ts}_{\bfone\lambda}$ is always sign-definite because $\PP^{(t)}_{\bfone}$ is proportional to the identity $\mathbbm{1}_{\text{in}}=\sum_\lambda \PP^{(s)}_\lambda$ on $\mathcal{H}_{\text{in}}=R_1\otimes R_2$. Because the recoupling matrix $M^{ts}_{\rho\lambda}$ performs a change of basis, it is invertible and hence all non-singlet rows must have at least one nonzero entry. For $M^{ts}_{\rho\lambda}$ to have vanishing dimension-weighted sum in Eq.~\eqref{eq:weightedzerosum}, a non-singlet row must contain at least two nonzero entries with opposite signs. This results in a non-singlet $\PP^{(t)}_\rho$ having indefinitely signed coefficients in the $s$-channel projector basis.

For $\mathrm{SO}(3)$ this statement is completely explicit. Consider the scattering process $\ell_1 + \ell_2\to\ell_2 + \ell_1$ of particles with isospins $\ell_1,\ell_2$. The incoming representations can couple to the isospins
\begin{equation}
  \ell_1\otimes\ell_2=\bigoplus_{J_s=|\ell_1-\ell_2|}^{\ell_1+\ell_2}J_s ,
\end{equation}
while the $t$-channel representations can couple to the intersection of the isospins in $\ell_1\otimes\ell_1$ and $\ell_2\otimes\ell_2$, i.e. $J=0,\dots,2\text{min}(\ell_1,\ell_2)$. The recoupling of a $t$-channel isospin $J$ projector into $s$-channel projectors, each given explicitly in App. \ref{app:C}, is
\begin{align}
  &\PP^{(t)}_J=\sum_{J_s=|\ell_1-\ell_2|}^{\ell_1+\ell_2}M^{ts}_{J J_s}\PP^{(s)}_{J_s},
  \label{eq:SO3recoupling}
  \\
  &M^{ts}_{J J_s}=(2J+1)(-1)^{J_s+\ell_1+\ell_2}
  \begin{Bmatrix}
    \ell_1&\ell_1&J\\
    \ell_2&\ell_2&J_s
  \end{Bmatrix} ,
  \label{eq:6j}
\end{align}
where $\big\{\begin{smallmatrix}
    j_1&j_2&j_3\\
    j_4&j_5&j_6
  \end{smallmatrix}\big\}$ is Wigner's $6j$-symbol. The singlet row is
\begin{equation}
  M^{ts}_{0J_s}={1\over\sqrt{(2\ell_1+1)(2\ell_2+1)} }>0,
  \label{eq:su2singletrow}
\end{equation}
whereas every $J>0$ row obeys
\begin{equation}
  \sum_{J_s=|\ell_1-\ell_2|}^{\ell_1+\ell_2}M^{ts}_{J J_s}(2J_s+1)=0 .
  \label{eq:su2zerosum}
\end{equation}

As a simple example showing the definite and indefinite signs for the singlet and non-singlets, respectively, consider the scattering of identical isospin 1 states. The $t$-channel projectors $\PP^{(t)}_0,\PP^{(t)}_1,\PP^{(t)}_2$ have the following expansion in the $s$-channel basis,
\begin{align}
    &\PP^{(t)}_0=\frac{1}{3}\PP^{(s)}_0+\frac{1}{3}\PP^{(s)}_1+\frac{1}{3}\PP^{(s)}_2\\
    &\PP^{(t)}_1=-\PP^{(s)}_0-\frac{1}{2}\PP^{(s)}_1+\frac{1}{2}\PP^{(s)}_2\\
    &\PP^{(t)}_2=\frac{5}{3}\PP^{(s)}_0-\frac{5}{6}\PP^{(s)}_1+\frac{1}{6}\PP^{(s)}_2
\end{align}

Because of this property, any nonzero time-delay coefficient carried by a $t$-channel non-singlet necessarily produces a negative physical delay eigenvalue in some incoming channel. 
This combined with the conclusion from Eq.~\eqref{eq:generaldelay} tells us that a non-singlet channel cannot have $\delta_\rho<1/2$ as claimed.

It is conceivable to consider whether some linear combination of $t$-channel non-singlet contributions $\chi _\rho$ with $\rho\neq\bfone$ to the time-delay can remove the negativity of the physical time-delays in some incoming channel that we discussed from a single non-singlet. Such a linear combination would contribute to the irreducible representation $\lambda$ physical time-delay as
\begin{align}
    \Delta T^{\text{ns}}_\lambda =\sum_{\rho\neq\bfone} M^{ts}_{\rho\lambda}\frac{\partial}{\partial E}\text{Re}\chi_\rho.
\end{align}
Without the $t$-channel singlet however, the $\Delta T^{\text{ns}}_\lambda$ are always sign indefinite. Indeed, because of Eq.~\eqref{eq:weightedzerosum} we have
\begin{align}
    \sum_\lambda\Delta T^{\text{ns}}_\lambda \text{dim}(\lambda)=0.
\end{align}
This implies that there must be a time-advance in some incoming channel, regardless of the linear combination, unless the the $t$-channel singlet is the dominant contribution.
\section{The weak-gravity condition}
\label{sec:weakgravity}

So far we have argued that a non-singlet intercept with $\alpha(0)>3/2$ leads to a sign-indefinite time-delay in the diffusion disk. This holds under the premise that the universal positive contribution from the graviton pole is sufficiently weak, due to its proportionality to $G_N$. Let us quantify how weak gravity must be for the constraint to hold.

Since Newton's constant is dimensionful, we measure gravity in units of the Regge scale, writing
\begin{equation}
  \lambda_{\rm grav}\equiv G_N M_\ast^{D-2},
  \label{eq:lambdagravdef}
\end{equation}
so that $\lambda_{\rm grav}\sim g_s^2$ in string theory. The gravitational time-delay is universal and positive in the incoming charge space, $\tdelay_{\rm grav}=G_N\,K_{\rm grav}(s,b)$, with
\begin{equation}
  K_{\rm grav}\sim \sqrt{s}\times
  \begin{cases}
    G_\perp(b), & b\gg b_{\diff,{\bfone}},\\[2pt]
    \alpha'_{\bfone}L\, D_{\bfone}(b,L), &
    R_S\ll b\lesssim b_{\diff,{\bfone}} .
  \end{cases}
  \label{eq:gravkernel}
\end{equation}
The useful quantity is then the critical gravitational coupling
\begin{equation}
 \lambda_{\rm grav}^{\rm crit}(s,b)
  \equiv M_\ast^{D-2}\,{m_B\,|\tdelay_B^{\rm ns}(s,b)|\over K_{\rm grav}(s,b)} ,
  \label{eq:GNcritdef}
\end{equation}
where $\tdelay_B^{\rm ns}$ is the non-singlet time-delay from the irreducible representation $B$ and $m_B\equiv \max_\lambda[-M^{ts}_{B\lambda}]_+$ is the largest negative entry in the non-singlet recoupling row. The graviton delay masks the dangerous non-singlet eigenvalue when $\lambda_{\rm grav}\gtrsim \lambda_{\rm grav}^{\rm crit}(s,b)$, so our weak-gravity condition corresponds to $\lambda_{\rm grav}\lesssim\lambda_{\rm grav}^{\rm crit}(s,b)$.

Inside the diffusion disk one finds
\begin{equation}
  \lambda_{{\rm grav},{\rm reg}}^{\rm crit,disk}
  \sim
 |1-\delta|
  \left({s\over M_*^2}\right)^{-\delta}
  {D_B(b,L)\over \alpha'_{\bfone}M_\ast^2 L\, D_{\bfone}(b,L)} ,
  \label{eq:lambdacritregdisk}
\end{equation}
and for comparable slopes $\alpha'$ with $b$ well inside both diffusion disks this reduces to
\begin{equation}
  \lambda_{{\rm grav},{\rm reg}}^{\rm crit,disk}
  \sim {1\over \alpha' L}
  \left({s\over M_*^2}\right)^{-\delta}.
  \label{eq:lambdacritregdisksimple}
\end{equation}
Eq.~\eqref{eq:lambdacritregdisksimple} sets the weak-gravity condition for non-singlet \emph{dominance} over the universal graviton time-delay.

The fact that gravity produces a definite positive contribution is reminiscent of how it enters in bounds on EFT coefficients in a gravitational theory~\cite{Caron-Huot:2021rmr}. There, some negativity of the $s^2$ EFT coefficient can be compensated for by the positive graviton pole contribution to dispersion relations. Any negativity of this Wilson coefficient larger than the graviton contribution is in conflict with unitarity or polynomial boundedness. Here we find that any non-singlet Regge trajectory which competes with the universal gravitational time-delay will violate causality, regardless of the sign.
\section{Conclusion}

We have argued that the eikonal phase for scattering with internal quantum numbers is an operator on the incoming charge space, and that causality constrains its physical $s$-channel eigenvalues. A $t$-channel singlet, such as the graviton pole, is proportional to the identity and gives a universal positive time-delay. By contrast, a non-singlet exchange recouples into the $s$-channel with both signs, so any nonzero non-singlet delay gives a time-advance in some incoming channel. For a Regge trajectory with $\alpha(0)=2-\delta$, this obstruction grows when $\delta<1/2$, and is marginal, with dimension-dependent behavior, at $\delta=1/2$. Since the compensating positive delay from the graviton is proportional to $G_N$, it can be made sufficiently small in the weak-gravity regime. Causality therefore requires that the leading Regge trajectory be an internal-symmetry singlet: non-singlet exchanges may be present, but they cannot control the leading high-energy behavior.

This result suggests several extensions. In AdS, the bulk eikonal phase is encoded in the Regge limit of boundary CFT correlators, so the absence of sign-indefinite time-advances should translate into constraints on charged Regge trajectories in large-$N$ CFTs with a weakly coupled gravitational dual. In particular, one expects the leading Regge singularity to be carried by the stress-tensor/graviton sector, while non-singlet trajectories should either have lower intercept or be accompanied by additional states that restore causal propagation. In de Sitter or inflationary settings, where an exact S-matrix is absent, an analogous constraint may appear in the high-energy, short-distance limit of cosmological correlators: sizable charge dependent leading Regge behavior would point to new degrees of freedom below the would-be cutoff, much as higher-derivative graviton interactions in \cite{Camanho:2014apa} require new higher-spin states to preserve causality.

\begin{acknowledgments}
We thank Calvin Y.-R. Chen, Kelian H{\"a}ring, Grant Remmen, Laurentiu Rodina and Justinas Rumbutis for discussions. The research of Y-t H is supported by Taiwan NSTC Grant
No. 112-2628-M-002-003-MY3 and L.W.L. is supported by the Taiwan NSTC Grant No. 113-2811-M-002-167-MY3 and the Yushan Young Fellowship.
\end{acknowledgments}

\appendix
\section{Non-singlet rows have indefinite sign}
\label{app:A}
The projection operators $\PP^{(s)}_\lambda$ and $\PP^{(t)}_\rho$, represented diagrammatically in Fig. \ref{fig:TtoS} are generally built from Clebsch--Gordan coefficients $C_{\rho}^{\lambda\nu}:V_\lambda\otimes V_\nu\to V_\rho$, which project vectors $v$ in the reducible tensor product space to the irreducible vector space, i.e. $(C_\rho^{\lambda\nu}\cdot v)\in V_\rho$ \cite{Birdtracks}. Importantly for the discussion, Clebsch--Gordan coefficients may be thought of as invariant tensors under the action $g\in G$,
\begin{align}
    (C_\rho^{\lambda\nu})_{I_\rho}^{\;J_{\lambda}J_{\nu}}\lambda(g)_{J_\lambda}^{\;I_\lambda}\nu(g)_{J_\nu}^{\;I_\nu}=\rho(g)_{I_\rho}^{\;J_\rho}(C_\rho^{\lambda\nu})_{J_\rho}^{\;I_{\lambda}I_{\nu}}
    \label{eq:Clebschinvariance}
\end{align}
where we have introduced vector space labels $I_\rho,I_\lambda,I_{\nu},$ etc., and repeated indices are summed over. 

The $s$-channel projection operators $\PP^{(s)}_\lambda:V_{R_1}\otimes V_{R_2}\to V_{\overline{R}_1}\otimes V_{\overline{R}_2}$ between two tensor product spaces, associated with an irreducible representation $\lambda$ is written in terms of the Clebsch--Gordan coefficients $C_{\lambda}^{R_1R_2}$ together with its conjugate $C^{\overline{\lambda}}_{\overline{R}_1\overline{R}_2}$,
\begin{align}
    \PP^{(s)}_\lambda=(C_{\lambda}^{R_1R_2})_{I_\lambda}(C^{\overline{\lambda}}_{\overline{R}_1\overline{R}_2})^{I_{\lambda}},
\end{align}
suppressing all other vector space indices. The $t$-channel projection operators $\PP^{(t)}_\rho:V_{R_1}\otimes V_{\overline{R}_1}\to V_{R_2}\otimes V_{\overline{R}_2}$ are constructed similarly,\footnote{As written in Eq.~\eqref{eq:Clebschtchannel}, $\PP^{(t)}_\rho$ are not Hermitian operators on $V_{R_1}\otimes V_{R_2}$ when $\rho$ is complex. To ensure the recoupling matrix $M^{ts}_{\rho\lambda}$ is real, we should change the $t$-channel basis of projectors to the Hermitian combinations $\PP^{(t)}_{\text{Re}\rho}=(\PP^{(t)}_\rho + \PP^{(t)}_{\overline\rho})/2$, $\PP^{(t)}_{\text{Im}\rho}=(\PP^{(t)}_\rho - \PP^{(t)}_{\overline\rho})/(2\mathrm{i})$.}
\begin{align}
    \PP^{(t)}_\rho=(C_{\rho}^{R_1\overline{R}_1})_{I_\rho}(C^{\overline{\rho}}_{R_2 \overline{R}_2})^{I_{\rho}}.
    \label{eq:Clebschtchannel}
\end{align}

Next, for finite dimensional unitary representations we have invariant Hermitian metrics $h_{I_{R_1}J_{\overline{R}_1}}$, $h_{I_{R_2}J_{\overline{R}_2}}$ which we use to define the $s$-channel trace
\begin{align}
    \text{tr}_s\left(A\right)\equiv h_{I_{R_1}J_{\overline{R}_1}}h_{I_{R_2}J_{\overline{R}_2}}A^{I_{R_1}I_{R_2};J_{\overline{R}_1}J_{\overline{R}_2}}.
\end{align}
We normalize the trace such that the $s$-channel trace of $\PP^{(s)}_\lambda$ is $\text{dim}(\lambda)$. Taking the $s$-channel trace of Eq.~\eqref{eq:Clebschtchannel}, we see that the metric $h_{I_{R_1}J_{\overline{R}_1}}$ contracts the indices of the Clebsch--Gordan coefficient $C_{\rho}^{R_1\overline{R}_1}$
\begin{align}
    X_{K_{\rho}}\equiv h_{I_{R_1}J_{\overline{R}_1}}(C_{\rho}^{R_1\overline{R}_1})_{K_\rho}^{\;I_{R_1}J_{\overline{R}_1}}
\end{align} 
Similarly, there is another vector $\overline{Y}_{I_\rho}$ generated by contracting $h_{I_{R_2}J_{\overline{R}_2}}$ with the other Clebsch--Gordan coefficient. The left-hand side of Fig. \ref{fig:ShursLemma} shows diagrammatically how the trace of Eq.~\eqref{eq:Clebschtchannel} is an invariant gluing of these two vectors $\text{tr}_s(\PP^{(t)}_{\rho})=X^{I_\rho}\overline{Y}_{I_{\rho}}$

Because of the invariance properties of the Clebsch--Gordan coefficients and metric, $X_{K_\rho}$ is also invariant under all $g\in G$,
\begin{align}
    \rho(g)_{I_\rho}^{\;J_\rho}X_{J_\rho}=X_{I_\rho}.
\end{align}
Schur's lemma implies among other things however that such a vector cannot exist for an irreducible representation unless $\rho$ is the trivial singlet representation. Thus we are led to the conclusion that $X_{I_{\rho}}=0$ for every non-singlet irreducible representation $\rho$. 

Because $\text{tr}_s(\PP^{(t)}_{\rho})=X^{I_\rho}\overline{Y}_{I_{\rho}}\propto\delta_{\rho\bfone}$, we get Eq.~\eqref{eq:weightedzerosum}. 
\section{$\PP^{(t)}_{\bfone}$ is proportional to the identity on the incoming Hilbert space}
\label{app:B}
That the $t$-channel singlet projector $\PP^{(t)}_\bfone$ is proportional to the identity $\mathbbm{1}_{\text{in}}$ in the incoming Hilbert space $V_{R_1}\otimes V_{R_2}$ is due to the fact that a vector $v_{I_{R_1}J_{\overline{R}_1}}\in V_{R_1}\otimes V_{\overline{R}_1}$ in the singlet representation is proportional to the invariant metric,
\begin{align}
    v_{I_{R_1}J_{\overline{R}_1}}= h_{I_{R_1}J_{\overline{R}_1}}v.
\end{align}
Because the $t$-channel singlet projector $\PP^{(t)}_\bfone$ projects onto singlet vectors in $V_{R_1}\otimes V_{\overline{R}_1}$ and converts them into singlets in $V_{R_2}\otimes V_{\overline{R}_2}$, it should be proportional to the product of the two metrics
\begin{align}
    (\PP^{(t)}_\bfone)_{I_{R_1}J_{\overline{R}_1};I_{R_2}J_{\overline{R}_2}}\propto h_{I_{R_1}J_{\overline{R}_1}}h_{I_{R_2}J_{\overline{R}_2}},
    \label{eq:Ptproptosinglet}
\end{align}
with all indices lowered.

The identity $\mathbbm{1}_{\text{in}}$ on $V_{R_1}\otimes V_{R_2}$ on the other hand should map any vector $v_{I_{R_1}I_{R_2}}$ to itself. Defining matrix multiplication with respect to the inverse metrics, it is straightforward to see that the right-hand side of Eq.~\eqref{eq:Ptproptosinglet} is the appropriate definition of $\mathbbm{1}_{\text{in}}$.
\section{Explicit $\mathrm{SO}(3)$ projectors}
\label{app:C}
We present here for concreteness the projectors for $\mathrm{SO}(3)$ defined in the tensor product space of isospin $\ell_1,\ell_2$ representations $\ell_1\otimes\ell_2$. There is a projector $\PP^{(s)}_J$ corresponding to each isospin $J=|\ell_1-\ell_2|,\dots,\ell_1+\ell_2$ making up the irreducible subspaces of $\ell_1\otimes\ell_2$,
\begin{align}
(\PP^{(s)}_J)&_{m_1m_2;m_3m_4}=(2J+1)(-1)^{\ell_1-m_1+\ell_2-m_2}\times\nonumber\\
&\times\sum_{M=-J}^J\begin{pmatrix}
        \ell_1 & \ell_2 & J\\
        m_1 & m_2 & -M
    \end{pmatrix}
    \begin{pmatrix}
        J & \ell_2 & \ell_1\\
        M & m_3 & m_4
    \end{pmatrix},
    \label{eq:SO3schannelprojector}
\end{align}
where $\big(\begin{smallmatrix}
        j_1 & j_2 & j_3\\
        m_1 & m_2 & m_3
    \end{smallmatrix}\big)$ are Wigner's standard $3j$-symbols. As written, Eq.~\eqref{eq:SO3schannelprojector} serves as projectors of $\mathrm{SO}(3)$ in the $s$-channel of the scattering process considered in the main text. The $t$-channel `projectors' act as operators $\PP^{(t)}_\lambda:V_{R_1}\otimes  V_{\overline{R}_1}\to V_{R_2}\otimes V_{\overline{R}_2}$, and for $\mathrm{SO}(3)$ can be obtained from Eq.~\eqref{eq:SO3schannelprojector} by consistently cyclically permuting the quantum numbers $(234)\to(423)$,
\begin{align}
(\PP^{(t)}_J)&_{m_1m_4;m_2m_3}=(2J+1)(-1)^{\ell_1-m_1+\ell_1-m_4}\times\nonumber\\
&\times\sum_{M=-J}^J\begin{pmatrix}
        \ell_1 & \ell_1 & J\\
        m_1 & m_4 & -M
    \end{pmatrix}
    \begin{pmatrix}
        J & \ell_2 & \ell_2\\
        M & m_2 & m_3
    \end{pmatrix},
    \label{eq:SO3tchannelprojector}
\end{align}
where $J=0,\dots,2\text{min}(\ell_1,\ell_2)$. Using these expressions, one can verify explicitly Eq.~\eqref{eq:SO3recoupling}.
\section{A heterotic string amplitude}
Heterotic string theory is a useful example with UV complete amplitudes to exhibit the interplay between causality and color, involving external states in the adjoint representation of the gauge group $G$, which we will take as $G=\mathrm{SO}(32)$, while also containing a graviton exchange. The tree level amplitude of four external gauge bosons in heterotic string theory is \cite{Gross}
\begin{align}
    A=&\,g^2\frac{\Gamma(-s/2)\Gamma(-t/2)\Gamma(-u/2)}{\Gamma(s/2)\Gamma(t/2)\Gamma(u/2)}K\times\nonumber\\
    &\Big[\Big(\frac{\text{tr}(T^{a_1}T^{a_2})\text{tr}(T^{a_3}T^{a_4})}{16s(1+s/2)}+\frac{15\text{tr}(T^{a_1}T^{a_2}T^{a_3}T^{a_4})}{st}\Big)\nonumber\\
    &+\text{cyclic perm's of }(234)\Big].
\end{align}
Here $K$ is the standard crossing symmetric kinematic prefactor present in supersymmetric string amplitudes encoding polarization data, scaling in momentum variables as $\sim p^4$. This amplitude is currently written in the trace basis, but can be converted to the $t$-channel projector basis $\PP^{(t)}_\lambda=(\PP^{(t)}_1,\dots,\PP^{(t)}_6)$, where $\PP^{(t)}_1$ is the singlet projector, and the rest correspond to the other representations two adjoint particles can couple to, with $\PP^{(t)}_5$ the adjoint projector, using the conventions of \cite{AYJJ}. 

The Regge behavior of this amplitude plainly shows the dominance of the $t$-channel singlet consistent with asymptotic causality, scaling as $\sim s^{2+t}\PP^{(t)}_1$, while the non-singlet representations have sub-dominant Regge behaviors with $\delta=1,2,3$. Explicitly, the Regge limit is
\begin{align}
    A\sim &\,\mathcal{E}g^2\Big(\frac{31}{1+t/2}s^{2+t}\PP^{(t)}_1+\frac{421}{4} ts^t\PP^{(t)}_2-\frac{29}{4}ts^t\PP^{(t)}_3\nonumber\\
    &+\frac{61}{4}ts^t\PP^{(t)}_4-225s^{1+t}\PP^{(t)}_5+\frac{1}{2}t(1-t/2)s^{-1+t}\PP^{(t)}_6\Big)\nonumber\\
    &\mathcal{E}=-(\epsilon_1\cdot\epsilon_4)(\epsilon_2\cdot\epsilon_3)\frac{\pi(\cot(\pi s/2)+\cot(\pi t/2))}{2^{t-1}\Gamma(1+t/2)^2},
\end{align}
where we assume for simplicity all polarizations are transverse to the momenta.

\begin{thebibliography}{0}%
\makeatletter
\providecommand \@ifxundefined [1]{%
 \@ifx{#1\undefined}
}%
\providecommand \@ifnum [1]{%
 \ifnum #1\expandafter \@firstoftwo
 \else \expandafter \@secondoftwo
 \fi
}%
\providecommand \@ifx [1]{%
 \ifx #1\expandafter \@firstoftwo
 \else \expandafter \@secondoftwo
 \fi
}%
\providecommand \natexlab [1]{#1}%
\providecommand \enquote  [1]{``#1''}%
\providecommand \bibnamefont  [1]{#1}%
\providecommand \bibfnamefont [1]{#1}%
\providecommand \citenamefont [1]{#1}%
\providecommand \href@noop [0]{\@secondoftwo}%
\providecommand \href [0]{\begingroup \@sanitize@url \@href}%
\providecommand \@href[1]{\@@startlink{#1}\@@href}%
\providecommand \@@href[1]{\endgroup#1\@@endlink}%
\providecommand \@sanitize@url [0]{\catcode `\\12\catcode `\$12\catcode `\&12\catcode `\#12\catcode `\^12\catcode `\_12\catcode `\%12\relax}%
\providecommand \@@startlink[1]{}%
\providecommand \@@endlink[0]{}%
\providecommand \url  [0]{\begingroup\@sanitize@url \@url }%
\providecommand \@url [1]{\endgroup\@href {#1}{\urlprefix }}%
\providecommand \urlprefix  [0]{URL }%
\providecommand \Eprint [0]{\href }%
\providecommand \doibase [0]{http://dx.doi.org/}%
\providecommand \selectlanguage [0]{\@gobble}%
\providecommand \bibinfo  [0]{\@secondoftwo}%
\providecommand \bibfield  [0]{\@secondoftwo}%
\providecommand \translation [1]{[#1]}%
\providecommand \BibitemOpen [0]{}%
\providecommand \bibitemStop [0]{}%
\providecommand \bibitemNoStop [0]{.\EOS\space}%
\providecommand \EOS [0]{\spacefactor3000\relax}%
\providecommand \BibitemShut  [1]{\csname bibitem#1\endcsname}%
\let\auto@bib@innerbib\@empty
\end{thebibliography}%


\begin{thebibliography}{99}

\bibitem{Weinberg:1965nx}
S.~Weinberg,
``Infrared photons and gravitons,''
Phys. Rev. \textbf{140}, B516--B524 (1965).


\bibitem{Calisto:2026ep}
F.~Calisto, C.~Cheung, G.~N.~Remmen, F.~Sciotti, and M.~Tarquini,
``The Equivalence Principle at High Energies Completes the Spectrum,''
arXiv:2605.20319 [hep-th].


\bibitem{Shapiro:1964uw}
I.~I.~Shapiro,
``Fourth test of general relativity,''
Phys. Rev. Lett. \textbf{13}, 789--791 (1964).





\bibitem{Camanho:2014apa}
X.~O.~Camanho, J.~D.~Edelstein, J.~Maldacena, and A.~Zhiboedov,
``Causality constraints on corrections to the graviton three-point coupling,''
JHEP \textbf{02}, 020 (2016),
arXiv:1407.5597 [hep-th].

\bibitem{deRham:2020zyh}
C.~de Rham and A.~J.~Tolley,
``Causality in Curved Spacetimes: The Speed of Light and Gravity,''
Phys. Rev. D {\bf 102}, no.8, 084048 (2020),
doi:10.1103/PhysRevD.102.084048
[arXiv:2007.01847 [hep-th]].

\bibitem{Chen:2021bvg}
C.~Y.~R.~Chen, C.~de Rham, A.~Margalit and A.~J.~Tolley,
``A cautionary case of casual causality,''
JHEP {\bf 03}, 025 (2022),
doi:10.1007/JHEP03(2022)025
[arXiv:2112.05031 [hep-th]].


\bibitem{Collins}
P.~D.~B.~Collins,
\emph{An Introduction to Regge Theory and High Energy Physics}
(Cambridge University Press, Cambridge, 1977).


\bibitem{ACV}
D.~Amati, M.~Ciafaloni, and G.~Veneziano,
Superstring collisions at Planckian energies,
Phys. Lett. B \textbf{197}, 81 (1987).


\bibitem{Haring:2022sdp}
K.~H\"aring and A.~Zhiboedov,
Gravitational Regge bounds,
SciPost Phys. \textbf{16}, 034 (2024),
\href{https://arxiv.org/abs/2202.08280}{arXiv:2202.08280}.
\bibitem{Caron-Huot:2021rmr}
S.~Caron-Huot, D.~Mazac, L.~Rastelli and D.~Simmons-Duffin,
JHEP \textbf{07}, 110 (2021)
doi:10.1007/JHEP07(2021)110

\bibitem{Wigner}
E.~P.~Wigner,
Lower limit for the energy derivative of the scattering phase shift,
Phys. Rev. \textbf{98}, 145 (1955).


\bibitem{Smith}
F.~T.~Smith,
Lifetime matrix in collision theory,
Phys. Rev. \textbf{118}, 349 (1960).

\bibitem{Martin:1976timeDelay}
P.~A.~Martin,
``On the time-delay of simple scattering systems,''
Commun. Math. Phys. {\bf 47}, 221--227 (1976),
doi:10.1007/BF01609841.
 
\bibitem{Birdtracks}
P.~Cvitanovic,
\textit{Group Theory: Birdtracks, Lie's, and Exceptional Groups},
(Princeton University Press, 2020).

\bibitem{Gross}
D.~J.~Gross and J.~A.~Harvey and E.~J.~Martinec and R.~Rohm,
Heterotic string theory: (II). The interacting heterotic string,
Nucl. Phys. B \textbf{267}, 75 (1986).

\bibitem{AYJJ}
A.~Hillman and Y.~Huang and L.~Rodina and J.~Rumbutis,
Spectral constraints on theories of colored particles and gravity,
Phys. Rev. Lett. \textbf{135}, 6 (2025).


\end{thebibliography}
\end{document}